\documentclass[12pt,preprint]{aastex}

\slugcomment{To appear in ApJ.}

\shorttitle{Keplerian Disk in L1527}
\shortauthors{Ohashi et al.}

\begin{document}


\title{Formation of a Keplerian disk in the infalling envelope around L1527 IRS: transformation from infalling motions to Kepler motions}


\author{Nagayoshi Ohashi\altaffilmark{1,2}, Kazuya Saigo\altaffilmark{3}, Yusuke Aso\altaffilmark{4}, Yuri Aikawa\altaffilmark{5}, Shin Koyamatsu\altaffilmark{4}, Masahiro N. Machida\altaffilmark{6}, Masao Saito\altaffilmark{7}, Sanemichi Z. Takahashi\altaffilmark{8}, Shigehisa Takakuwa\altaffilmark{2}, Kengo Tomida\altaffilmark{9,10}, Kohji Tomisaka\altaffilmark{11}, and Hsi-Wei Yen\altaffilmark{2}}





\altaffiltext{1}{Subaru Telescope, National Astronomical Observatory of Japan, 650 North A'ohoku Place, Hilo, HI 96720, USA; nohashi@naoj.org}
\altaffiltext{2}{Academia Sinica Institute of Astronomy and Astrophysics, P.O. Box 23-141, Taipei 10617, Taiwan}
\altaffiltext{3}{Chile Observatory, National Astronomical Observatory of Japan, Osawa 2-21-1, Mitaka, Tokyo 181-8588, Japan}
\altaffiltext{4}{Department of Astronomy, Graduate School of Science, The University of Tokyo, 7-3-1 Hongo, Bunkyo-ku, Tokyo 113-0033, Japan}
\altaffiltext{5}{Department of Earth and Planetary Sciences, Kobe University, Kobe 657-8501, Japan}
\altaffiltext{6}{Department of Earth and Planetary Sciences, Faculty of Sciences, Kyushu University, Fukuoka 812-8581, Japan}
\altaffiltext{7}{Joint ALMA Observatory, Ave. Alonso de Cordova 3107, Vitacura, Santiago, Chile}
\altaffiltext{8}{Department of Physics, Kyoto University, Oiwake-cho, Kitashirakawa, Sakyo-ku, Kyoto 606-8502, Japan}
\altaffiltext{9}{Department of Astronomical Science, Princeton University, Princeton, NJ 08544, USA}
\altaffiltext{10}{Department of Physics, The University of Tokyo, 7-3-1 Hongo, Bunkyo-ku, Tokyo 113-0033, Japan}
\altaffiltext{11}{National Astronomical Observatory of Japan, Osawa 2-21-1, Mitaka, Tokyo 181-8588, Japan}


\begin{abstract}
We report Atacama Large Millimeter/submillimeter Array (ALMA) cycle 0 observations of C$^{18}$O ($J=2-1$), SO ($J _N= 6_5-5_4$) and 1.3~mm dust continuum toward L1527 IRS, a class 0 solar-type protostar surrounded by an infalling and rotating envelope. C$^{18}$O emission shows strong redshifted absorption against the bright continuum emission associated with L1527 IRS, strongly suggesting infall motions in the C$^{18}$O envelope. The C$^{18}$O envelope also rotates with a velocity mostly proportional to $r^{-1}$, where $r$ is the radius, while the rotation profile at the innermost radius ($\sim$54~AU) may be shallower than $r^{-1}$, suggestive of formation of a Keplerian disk around the central protostar of 
$\sim0.3~M_{\sun}$ in dynamical mass. 
SO emission arising from the inner part of the C$^{18}$O envelope also shows rotation in the same direction as the C$^{18}$O envelope. The rotation is, however, rigid-body like which is very different from the differential rotation shown by C$^{18}$O.
In order to explain the line profiles and the position-velocity (PV) diagrams of C$^{18}$O and SO observed,
simple models composed of an infalling envelope surrounding a Keplerian disk of 54~AU in radius orbiting a star of 0.3~$M_{\sun}$ are examined.
It is found that 
in order to reproduce characteristic features of the observed line profiles and PV diagrams,
the infall velocity in the model has to be smaller than the free-fall velocity yielded by a star of 0.3~$M_{\sun}$.
Possible reasons for the reduced infall velocities are discussed.

\end{abstract}





\section{Introduction}
Keplerian disks are considered to be universally formed in the course of low-mass star formation, and play an essential role in the planet formation. In fact, a large fraction of low-mass pre-main-sequence stars, T Tauri stars are associated with disks \citep[e.g.,][]{beckwith90,andrews05,andrews07} , and several of these disks are confirmed to have Kepler rotation \citep[e.g.,][]{dutrey98,simon00}. Its formation process is, however, still poorly understood. In the conventional picture of star formation, a Keplerian disk is formed in a slowly rotating dense molecular core accreting onto a central protostar because of angular momentum conservation of the infalling material \citep{terebey84}. On the other hand, recent theoretical simulations suggest that magnetic filed must have strong impact on the formation of Keplerian disks around protostars \citep{hennebelle09,machida11}; in some cases with strong magnetic field Keplerian disks cannot be formed because of the magnetic braking of rotation \citep{mellon08, li11} although there are some controversial issues \citep{joos12,machida13,krumholz13}. While recent observations start showing kinematical evidence for Keplerian disks around low-mass protostars, in particular class I sources \citep{takakuwa12,yen13,murillo13,harsono14,yen14}, it is still extremely crucial for us to observationally investigate how Keplerian disks are formed around protostars, particularly 
those in the earliest evolutional stage.

L1527 IRS (hereafter L1527), originally found as an infrared point source (IRAS~04368+2557) by the Infrared Astronomical Satellite (IRAS)\citep{beichman86}, is located in one of the closest low-mass star forming region, Taurus Molecular Cloud at a distance of about 140 pc, and has been considered to be in the earliest evolutional stage (class 0) of star formation. The systemic velocity of L1527 was  measured to be $\sim$5.7~km~s$^{-1}$ from C$^{18}$O $J=1-0$ observations using the Nobeyama 45~m single-dish telescope, while it was measured to be $\sim$5.9~km~s$^{-1}$ from N$_2$H$^+$ ($J=1-0$) single-dish observations \citep{caselli02,tobin11}. We adopt 5.9~km~s$^{-1}$ for the systemic velocity of the L1527 system in this paper.
The first detailed observations of L1527 at a high angular resolution were made by Nobeyama Millimeter Array (NMA), revealing a flattened, edge-on envelope of 2000 AU in radius, perpendicular to the direction of the molecular outflow ejecting in the east-west direction \citep{ohashi97}. More importantly, they have suggested that the flattened envelope has dynamical infalling motions with slow rotation. The mass infalling rate was estimated to be 1$\times10^{-6}~M_{\sun}$~yr$^{-1}$. Following up observations at even higher angular resolutions were carried out by the Submillimeter Array (SMA) in C$^{18}$O ($J=2-1$) and the Combined Array for Research in Millimeter-wave Astronomy (CARMA) in $^{13}$CO ($J=2-1$). These observations have revealed that the rotation profile of the infalling envelope follows the negative first power of the radius up to $\sim$140 AU in radius \citep{yen13}, while the power may become -0.5 within a radius of $\sim$140 AU where a Keplerian disk might exist \citep{tobin12}. These previous observations, however, had relatively poor sensitivity to study kinematics of the envelope and the possible Keplerian disk more accurately. In this Paper, we report Atacama Large Millimeter/submillimeter Array (ALMA) cycle 0 observations of the protostar L1527, in C$^{18}$O ($J=2-1$), SO ($J _N= 6_5-5_4$) and 1.3~mm dust continuum. We note that C$^{18}$O is more likely optically thin, and thus a reliable tracer to investigate kinematics of the innermost envelope, where a Keplerian disk would be formed, as compared with $^{13}$CO.

\section{Observations}


ALMA having enormously high sensitivity as compared with previous interferometers was used to observe L1527 in C$^{18}$O $2-1$, SO $6_5-5_4$, and 1.3~mm dust continuum emissions on August 26, 2012. The total number of the 12 m antenna was 25, and the total on-source time was $\sim$37~minutes. The velocity resolution of the line emission was $\sim0.17$~km~s$^{-1}$. The maximum baseline was 366~m, providing an angular resolution of $0\farcs96\times0\farcs73$ in C$^{18}$O ($0\farcs8$ in geometrical mean), while the minimum baseline was 18~m, setting the maximum detectable size at $\sim9\arcsec$ for 30\% level detection or $\sim7\arcsec$ for 50\% level detection. The rms noise levels (1~$\sigma$) of C$^{18}$O and SO are 8.0~mJy~beam$^{-1}$, corresponding to 0.29~K in brightness temperature, and 9.5~mJy~beam$^{-1}$, corresponding to 0.35~K, respectively with the velocity resolution of 0.17~km~s$^{-1}$. The noise level of the continuum map is $\sim$5.3~mJy. All the observation parameters are summarized in Table~1.




\section{Results}
\subsection{1.3 mm continuum emission}
Compact emission of 1.31~mm continuum having a peak of $\sim$130 mJy~beam$^{-1}$ was detected.  The peak position measured with a Gaussian fitting is consistent with a previous measurement at high angular resolutions (Tobin et al. 2012, Yen et al. 2013). The continuum emission was barely resolved at the current angular resolution, providing a deconvolved size of $0\farcs64\times0\farcs36$ (PA$\sim0.9\degr$), corresponding to $\sim$90~AU$\times$50~AU at the distance of 140~pc. Total flux density integrated over the area with more than 3~$\sigma$ level is measured to be $\sim$0.20~Jy. We use the continuum peak position as the protostellar position of L1527 throughout this paper.

\subsection{C$^{18}$O emission}
C$^{18}$O $2-1$ was detected at more than 6 $\sigma$ level from $V_{\rm LSR}=3.1~$km~s$^{-1}$ (or -2.7 km/s in $\Delta V$ with respect to the systemic velocity of 5.9~km~s$^{-1}$)  to 8.5 km~s$^{-1}$ ($\Delta V=$2.6 km~s$^{-1}$) in LSR velocity, except for $V_{\rm LSR}=$6.0~km~s$^{-1}$ where C$^{18}$O shows absorption. The C$^{18}$O 2-1 line profile at the position of the central protostar is shown in Fig.~1a.

As was mentioned above, the C$^{18}$O $2-1$ emission shows a very strong, negative dip at $V_{\rm LSR}=$6.0~km~s$^{-1}$ and 6.1~km~s$^{-1}$, which are slightly redshifted as compared with the systemic velocity of 5.9~km~s$^{-1}$. The negative dip reaches $\sim-2.5$~K, corresponding to $\sim9~\sigma$, and should mostly be produced by absorption because the resolving out effect cannot make it.
The position and the size of the negative dip seen in the channel map are basically the same as those of the continuum emission, strongly suggesting that the negative dip is absorption against the continuum emission. Such a negative redshifted absorption, or the so-called inverse P-Cygni profile, is naturally explained by infall motions as previous studies discussed \citep{francesco01,pineda12}, including \citet{kristensen12} who actually observed an inverse P-Cygni profile toward L1527 in H$_2$O $1_{10}-1_{01}$. The other remarkable feature of the C$^{18}$O 2-1 line profile is two "shoulders", or plateaus at $V_{\rm LSR} =$4.5~km~s$^{-1}$ and 7.5~km~s$^{-1}$. 

The moment 0 map of C$^{18}$O $2-1$ integrated over $V_{\rm LSR}$ between 3.1~km~s$^{-1}$ and 8.5~km~s$^{-1}$ shows a flattened structure elongated in the north-south direction (Fig.~2a). 
This flattened structure, which is consistent with those observed previously in L1527 \citep{ohashi97, tobin12, yen13}, is basically the envelope surrounding L1527 IRS, as the previous studies also interpreted. The intensity-weighted mean velocity (i.e., moment ~1) map shows a clear velocity gradient from the south (blueshifted) to the north (redshifted), in the same direction of the elongation of the flattened structure. This velocity gradient, which has been also detected in previous observations \citep{ohashi97,tobin12,yen13}, is naturally considered to be due to rotation of the flattened structure.
The position-velocity diagram cutting through the line perpendicular to the rotation axis (Fig.~3a)
clearly demonstrates that the rotation is differential with a larger angular velocity at a smaller radius.



The total integrated intensity of C$^{18}$O is measured to be 2.7~Jy~km~s$^{-1}$, from which the total H$_2$ gas mass of the C$^{18}$O  envelope is estimated to be $1.1\times10^{-3}~M_{\sun}$, on the assumption that the line is optically thin, the C$^{18}$O  excitation temperature is 10~K, and the the relative abundance of C$^{18}$O is $3\times10^{-7}$. Since the excitation temperature assumed here would be a lower limit to the actual value, the total H$_2$ mass estimated should be considered as an upper limit to the actual gas mass.  

In summary, 
the C$^{18}$O emission traces an flattened envelope having both infall and rotation motions. Details of the kinematics will be discussed in  \S4.

\subsection{SO emission}
In addition to C$^{18}$O $2-1$ emission, SO emission was significantly detected in the velocity range of 4.0~km~s$^{-1}$ to 7.7~km~s$^{-1}$ in the LSR velocity at more than 6 $\sigma$ level. The line profile of SO measured at the position of the central protostar, shown in Fig.~1b, has a single peak at a velocity close to the systemic velocity, which is quite different from the C$^{18}$O profile having a deep absorption feature at a velocity close to the systemic velocity. The map of SO integrated over $V_{\rm LSR}$ between 4.0~km~s$^{-1}$ and 7.7~km~s$^{-1}$ shows a compact structure, located in the inner region of the C$^{18}$O flattened envelope (Fig.~2b). The deconvolved size of the SO emission was measured to be $1\farcs4\times0\farcs68$, corresponding to 200~AU$\times$95~AU, with a Gaussian fitting to the moment~0 map. 

The moment~1 map of SO shows a clear velocity gradient in the north-south direction, where C$^{18}$O emission also has a similar velocity gradient. Since the north-south velocity gradient of the C$^{18}$O emission is due to rotation, it is naturally considered that the north-south velocity gradient of the SO emission is also due to rotation. However, the PV diagram of the SO emission along the north-south direction, presented in Fig.~3b, shows velocity structures that are quite different from those seen in the C$^{18}$O PV diagram shown in Fig.~3a; between $\sim-1$~km~s$^{-1}$ to $\sim1$~ km~s$^{-1}$ in the relative velocity (or between $\sim5$~km~s$^{-1}$ to $\sim7$~ km~s$^{-1}$ in the LSR velocity), the velocity changes almost linearly as a function of the position, as shown in the green line in Fig.~3b. In addition to the component showing a linear velocity gradient, there would be additional components at higher velocities than $\sim$1~km~s$^{-1}$ in the relative velocity close to the central star. 

Since the SO emission is spatially compact as compared with the C$^{18}$O emission, it would be naturally considered that the SO emission arises from the inside of the C$^{18}$O flattened envelope. The fact that the kinematical nature of the SO emission shown in its line profile and PV diagram is quite different from that of the C$^{18}$O emission may suggest that the SO emission traces some particular regions within the C$^{18}$O envelope. The nature of the SO emission will be discussed in \S4.2.


\section{Discussions}
\subsection{Rotation of the C$^{18}$O flattened envelope and possible formation of a Keplerian Disk}
As was discussed in the previous section, the C$^{18}$O flattened envelope shows a clear velocity gradient along the direction of the elongation, which can be interpreted as rotation. The PV diagram in Fig.~3a shows that the rotation is differential having higher rotation velocities as the position gets closer to the central star. In order to understand the nature of the rotation, the power-law dependence of the rotation profile derived from the PV diagram is investigated in detail in this subsection.

The method to examine the power-law dependence of the rotation profile is the same as that described in Yen et al. (2013); while Yen et al. used another PV diagram covering larger scale structures to measure rotation velocities at larger radii $\gtrsim$500~AU, we do not use it. The PV diagram shows its data points measured on the diagram, and these data points are plotted with logarithmic scale in Fig.~4a. For a comparison purpose, previous measurements of the SMA observations \citep{yen13} were also plotted together.
From the comparison it is found that the ALMA and SMA measurements at radius more than 100~AU show similar slopes,
although the ALMA measurements provide higher rotation velocities than the SMA measurements. Rotation velocities measured by ALMA at smaller radii ($\leq$100~AU) are symmetric between blueshifted and redshifted velocities, while those measured at larger radii ($\geq$100~AU) are not symmetric with respect to the systemic velocity, suggesting a potential problem of the measurements at larger radii probably due to resolving out larger scale structures by ALMA (see more details below). For smaller radii, the systemic velocity of 5.9~km~s$^{-1}$ seems to be quite reasonable even though it was measured by single-dish observations.

A least-square fitting using power-law ($\propto r^{\alpha}$) to the all the data points including both SMA and ALMA measurements are performed, finding that the overall rotation profile can be fitted with a power-law having an index of $\sim-0.7$. This value is larger than that measured by SMA on similar scales. However, 50\% of emission arising from the structures having a $\sim$1000~AU scale (500 AU in radius) is missed in our current ALMA measurements (see \S2), and as a result, a missing flux is larger for emission at lower rotation velocities on larger scales. In such a situation, rotation velocities on larger scales are overestimated. For this reason, we focus on the measurements at radii less than 100~AU, shown in Fig.~4b, in the following discussions.

Another least-square fitting is performed to the ALMA data measured within 100~AU in radius, finding that the rotation profile at the radius less than 100 AU can be fitted with a power-law having an index of $\sim-1$, as shown in the dashed line in Fig.~4b. The sum of squared residuals, 
$\sum(\log V_{\rm obs}-\log V_{\rm model})^2/n$, where $n$ is degrees of freedom, is $2.6\times10^{-4}$. As was shown by \citet{yen13}, the power-law of the rotation profile at the radius larger than 100~AU is also $\sim-1$, suggesting that a similar rotation profile seems to continue even beyond a radius of 100 AU. \citet{tobin12} showed that the power-law of the rotation around L1527 is $-0.5$ at radius less than 100 AU, i.e., Kepler rotation, which is inconsistent with our result. For comparison purpose, the rotation curve obtained by \citet{tobin12} based on the best fitting to their measurements is also shown in Fig.~4b in the dotted line, apparently demonstrating that their rotation curve is inconsistent with ours. Since \citet{tobin12} have assumed Kepler rotation in their fitting, it is not clear whether or not their data can be explained better by a rotation curve with a power-law index of $\sim-1$. These results suggest that the rotation of the flattened envelope around L1527 can be explained as a case where angular momentum is conserved most probably because the material in the envelope is still infalling. 

A very close inspection of the ALMA measurements shown in Fig.~4b found that the rotation profile could become slightly shallower at the innermost rotation radius. In order for us to investigate this possibility, the measured data points within 100~AU are fitted with two power-laws having a break at a certain radius. The best fitting result shown in the dashed line in Fig.~4b demonstrates that the rotation profile within 100~AU could be explained by two power-laws having a break at a radius of $53.7\pm0.4$~AU. The sum of squared residuals is $1.3\times10^{-4}$, suggesting that the fitting with two power-laws is somewhat better than that with the single power-law even when smaller degrees of freedom with two power-laws is taken into account. The shallower rotation profile within $\sim$54 AU has an power-law index of $\sim-0.41\pm0.24$, whereas the index outside $\sim$54~AU is $\sim-1.16\pm0.13$.  Although the power-law within the break is  $\sim-0.4$ with a relatively larger error, this could suggest existence of a Keplerian disk of $\sim$50~AU in radius. If this is the case, the dynamical mass of the central protostar is estimated to be $\sim0.3~M_{\sun}$. We note that if the break due to formation of a Keplerian disk appears at a smaller radius than 50~AU on the dashed line in Fig. 4b, the resultant dynamical mass becomes larger than $0.3~M_{\sun}$. In this sense, $\sim0.3~M_{\sun}$ can be considered as a lower limit to the dynamical mass of the central protostar.

It is important to estimate the mass of the possible Keplerian disk. Although the size of the continuum emission, $\sim$90~AU$\times$50~AU (see Sec. 4.1) is slightly larger than the size of the disk, it would be still feasible to use the continuum emission to estimate the mass of the disk. On the assumption that the dust mass opacity is 0.1~cm$^2$~g$^{-1}$ at 220~GHz, which is extrapolated from 0.1~cm$^2$~g$^{-1}$ \citep{beckwith90} at 1~THz with $\beta$=0 \citep{tobin13}, and the dust temperature is $\sim$50~K, which is estimated from the temperature profile (see Sec. 4.3), the mass of the disk is estimated to be $\sim 2.8 \times 10^{-3}~M_\sun$. When the mass opacity at 220~GHz is assumed to be 0.022~~cm$^2$~g$^{-1}$, which is derived with $\beta=1$, the mass is estimated to be $\sim 1.3 \times 10^{-2}~M_\sun$. Since the size of the dust emission is slightly larger than the size of the disk, these masses may give upper limits to the actual disk mass. Although the disk mass still strongly depends on the dust mass opacity, it seems to be at least one order of magnitude smaller than the mass of the central protostar, suggesting that the disk would be gravitationally stable.

\subsection{Origin of the SO emission: a ring?}
As shown in the \S3.3, the SO emission arising from the inside of the C$^{18}$O flattened envelope also shows rotation although the nature of the rotation of the SO emission is quite different from that of the C$^{18}$O emission. How these two emission lines show quite different kinematics even though they arise from similar regions?

One possible and natural explanation is that only a part of the whole envelope is seen in SO. Suppose the SO emission arises from a rotating ring having an edge-on configuration to observers. Such a ring rotates at a constant angular velocity, while the observed velocity gradient due to the rotation becomes linear because of a projection effect.


The linear velocity gradient is seen between $\sim-1$~km~s$^{-1}$ to $\sim1$~ km~s$^{-1}$ in the relative velocity, suggesting that the rotation velocity of the ring is $\sim1$~ km~s$^{-1}$. Since this ring is a part of the C$^{18}$O flattened envelope, the radius of the ring can be estimated to be $\sim$120~AU according to the rotation profile of the C$^{18}$O emission. The radius of the ring is
larger than that of the possible Keplerian disk discussed in the previous section, suggesting that the ring is located in the infalling envelope. On the other hand, the ring has probably some radial width because there is additional SO emission at higher velocities than  $\sim1$~ km~s$^{-1}$ in the relative velocity, located closer to the central star.

How can only a part of the whole envelope be seen in SO as a ring?
The most probable solution to make it possible is to locally enhance the fractional abundance of SO in the ring.
SO molecules in envelopes are considered to be depleted on dust particles because of its low temperature ($\sim$10 K), whereas they can come back
to gas phase when the temperature of dust particles becomes high enough; 
when the adsorption energy of SO is assumed to be 2600 K based on laboratory data \citep{garrod06}, its sublimation temperature is estimated to be $\sim$60~K, while it is estimated to be $\sim$40~K assuming that the adsorption energy is 2000~K \citep{hasegawa93}. Although the sublimation temperature of SO has some uncertainty, this suggests that when the dust temperature becomes higher than $\sim$40-60~K, the abundance of gaseous SO becomes larger.

If the radius of the ring is $\sim$120~AU as was mentioned above, the dust temperature
has to become locally higher than $\sim$40-60 K at a radius of $\sim$120~AU.
Such a high dust temperature cannot happen at a radius of $\sim$120~AU 
with the radiation from the central protostar; the radiative equilibrium temperature at a radius of 120~AU around L1527 is estimated to be $\sim$33~K (see Sec. 4.3).
When accreting material releases its kinetic energy as accretion shock, dust temperature
could become higher. Accreting material around L1527 has its accreting velocity of $\sim$2~km~s$^{-1}$ 
at a radius of $\sim$120~AU with a stellar mass of  $\sim0.3~M_{\sun}$, which was estimated in \S 4.1. With this accreting velocity, together with a H$_2$ density of $\sim$10$^7$~ cm$^{-3}$ (see Sec. 4.3, and also Fig. 6), the dust temperature may become $\sim$40~K \citep{neufeld94,aota14}.
We note that the radius of the hot ring where accretion shock may occur is larger than the radius of the possible Keplerian disk discussed in the previous section. Such a case can be explained when we consider that the streamlines of the flow upstream of the shock are well approximated by ballistic trajectories\citep{cassen81}.


Although dust temperature is expected to become higher due to a shock, it is still not clear what the actual dust temperature at a radius of 120~AU. Since it is not easy to estimate the dust temperature from the current observational data, we estimate the kinetic temperature of the SO gas instead of the dust temperature. In order to estimate
the kinetic temperature of the SO gas, we perform statistical equilibrium calculations based on
large velocity gradient (LVG) model\citep{goldreich74}. Details of the calculations are described in Appendix.
According to the calculations, it is found that the kinetic temperature of the SO gas is most likely $\sim$32~K, which is not as high as 40-60~K, but quite similar to the radiative equilibrium temperature at a radius of 120~AU around L1527 of 2.7 L$_\sun$ in luminosity. Nevertheless, without a hot region, the abundance of the SO molecule cannot be locally enhanced at a radius of 120~AU. We therefore suggest that the SO ring is mostly $\sim$32~K, but has a thin layer with a higher temperature at its most outer radius. 

Such a ring with two zones of different temperatures is actually quite reasonable because
the SO gas in the post-shock region may cool down very quickly and its temperature becomes the ambient temperature. According to theoretical calculations\citep{aota14}, a high temperature zone in a post-shock region is spatially very thin. Since the LVG calculations assume a structure with one zone, it is impossible for us to obtain two temperatures. The reason why the temperature estimated from the LVG calculations is 32~K instead of a higher temperature is that most of the SO emission arises from the 32~K zone, which is optically thick, rather than the high temperature zone, which is spatially and optically thin. 

One may wonder how long SO molecules can stay in gas phase after the temperature becomes below the sublimation temperature. Its time scale is given by $10^3\times(10^7/n_{{\rm H}}~{\rm cm}^{-3})$~yr. Since $n_{{\rm H}_2}$ at 120~AU is $\sim10^7$~cm$^{-3}$ (see Sec. 4.3; see also Fig.~6), the SO molecules can stay in gas phase for $\sim500$~yr. Since SO gas together with H$_2$ gas in the envelope accretes toward the central star at a speed of $\sim$0.5~km~s$^{-1}$ at 120~AU in radius (see Sec. 4.3), the SO ring can spreads inwards across a width of $\sim$50~AU.
In the next section, models of an infalling envelope including a possible SO ring will be used to reproduce the observations. Note that if a different infall velocity is assumed, the width of the SO ring also changes although the width is fixed at 50~AU in our models (see Sec 4.3).

Higher transition of SO was also detected in L1527\citep{sakai14}, which also has suggested that the SO emission may be due to shock. SO emission having similar nature was also detected in L1489 IRS \citep{yen14}.

\subsection{Infalling motions in the flattened envelope}
As was shown above, the C$^{18}$O line profile shows a negative absorption at the systemic velocity. This redshifted absorption is created against the continuum emission, and should be due to infalling motions in the envelope. As was discussed in the previous subsection, the overall rotation profile of the C$^{18}$O flattened envelope can be explained with a power-law having an index of $\sim-1$, suggesting conservation of the specific angular momentum of the rotating material. Such a case is naturally expected when the rotating material is also infalling at the same time. In this subsection, nature of the infalling motions as well as rotation in the flattened envelope is discussed in more detail.

In order for us to investigate nature of the motions in the envelope, we construct models to reproduce C$^{18}$O  and SO line profiles and PV diagrams. As shown in Fig. 5, our models are composed of three parts, i.e., an envelope, an SO ring, and a Keplerian disk. 
The model envelope is based on that discussed by \citet{tobin08}, who modified the infalling envelope model of  \citet{terebey84} to reproduce the NIR scattering light around L1527. Note that the structure of the model envelope is not flattened even though the observed envelope has a flattened structure. It is, however, still reasonable to use such a model envelope because we basically discuss only the line profile at the center and the PV diagram cutting along the disk mid-plane. The model envelope rotates with a power-law index of -1 ($j=6.1\times10^{-4}$~km~s$^{-1}$~pc), while it has a temperature profile with a radial dependence of $360(r/1 \rm{AU})^{-0.5}$, which is the radiative equilibrium temperature in Kelvin with the total luminosity of L1527 (2.75~$L_{\sun}$; \citet{tobin08}) under optically thin assumption. The Model envelope has a SO ring region of 32~K in temperature with its outer and inner radii of 120~AU and 70~AU, respectively. Within the SO ring region, the fractional abundance of SO is $9\times10^{-7}$, which is 5000 times higher than that outside and inside of the hot ring. Within 54~AU in radius, a Keplerian disk orbiting a central star of 0.3~$M_{\sun}$ is considered, based on the power-law disk model\citep[e.g.][]{kitamura02}. Infall velocities in model envelopes are adjusted to reproduce the observations, as will be described below. Expected line emissions along line of sights were calculated with radiative transfer equations assuming the local thermal equilibrium and an edge-on configuration of the ring and the disk (their inclination angles are 90$\degr$), and the CASA simulator was used to reproduce the observed line profiles and PV diagrams. 

In the first model (Model~1), infall velocities are simply equal to the free-fall velocities 
yielded by the gravity of the central star, while there is no infall in the Keplerian disk,
as shown in Fig. 6a.
The C$^{18}$O line profiles produced 
with this model is shown in Fig. 1a in the red line. The C$^{18}$O profile of Model~1 shows a typical infall profile having a deep dip at a redshifted velocity and a stronger blueshifted peak, which is similar to the observed profile. However,  the velocities of the two peaks in the model profile are not consistent with those in the observations. This suggests that the infall velocity in the model is too large because two peaks in an infalling profile appear at the characteristic infall velocity. 

The C$^{18}$O PV diagram of Model~1 is compared with the observation 
in Fig. 3c. Although the overall shapes of the PV diagram produced from the model are similar to
the observations, the model cannot reproduce emissions at lower blueshifted and redshifted velocities ($\Delta$V$\leq$0.5~km~s$^{-1}$) well. In addition, the blueshifted emission between 
$V_{\rm LSR}$=4.5 and 5.5~km~s$^{-1}$ in the model PV diagram is more extended to the north compared with the observation. 
Similarly, the redshifted emission  between $V_{\rm LSR}$=7~km~s$^{-1}$ and 7.5~km~s$^{-1}$ in the model PV diagram extends more to 
the south compared with the observation. The discrepancies are also considered to be due to the larger 
infalling velocity making two peaks at higher velocities in the model line profile shown in Fig. 1c.

The SO line profile produced by Model~1 presented in Fig. 1b shows
that the model profile has two strong peaks at V$_{\rm LSR}\sim$4~km~s$^{-1}$ and 
$\sim$8~km~s$^{-1}$, which is
very different from the observations having a single peak at the systemic velocity of 5.9~km~s$^{-1}$.
Because of the strong two peaks, the line profile is much wider than the observation as well.
The SO PV diagram produced by Model 1 is also compared with the observation in Fig. 3d;
the model PV diagram has two strong peaks at V$_{\rm LSR}\sim$4~km~s$^{-1}$
and $\sim$8~km~s$^{-1}$,
while the observed PV diagram has peaks close to the systemic velocity.
In addition to the discrepancy in the peak velocities, the overall velocity gradient of the SO emission
in the model PV diagram is larger than that in the observed PV diagram.
These two strong peaks at higher velocities are due to the infall motions, suggesting that the infall velocity is too high to reproduce the observations.

In order to reduce the discrepancies described above, 
another model (Model~2) is made with a different radial profile of the infall velocity, as shown in Fig. 6b.
The infalling velocity at radii $\geq$250~AU is a half of the free-fall velocity yielded by the gravity of the central star. This makes the characteristic infall velocity of the C$^{18}$O envelope slower.
In addition, the infall velocity between 120~AU and 54~AU in radius is slower than the free-fall velocity by 1/5. This makes the infall velocity in the SO ring significantly slower.
The infalling velocity at radii between 54~ AU and 250~AU except for the hot ring region remains the free-fall velocity possibly because the infalling gas is decoupled from the magnetic field by ambipolar diffusion\citep{mellon09}.

The resultant C$^{18}$O profile calculated from Model 2 is shown in Fig. 1c in the red line; it has two peaks 
at 5.4~km~s$^{-1}$ and 6.5~km~s$^{-1}$, which is consistent with the observation. The model profile also shows two additional peaks at higher blueshifted and redshifted velocities where the observed profile has the shoulders. Although these additional peaks do not appear as shoulders like observations, the overall shape of the observed C$^{18}$O line profile is mostly reproduced by
that calculated from Model 2. Note that the reason why the two additional peaks appear in the model profile is that the infall velocity remains the free-fall velocities towards a 0.3~$M_{\sun}$ protostar at radii between 120~AU and 250~AU. 
The resultant C$^{18}$O PV diagram of Model 2 is compared with the observed PV
diagram in Fig. 3e; Model 2 reproduces the overall characteristics of the observed PV diagram, including lower blueshifted and redshifted velocities that cannot be reproduced by Model 1 well.
The diagram made from Model~2, on the other hand, shows dips at V$_{\rm LSR}\sim$5~km~s$^{-1}$
and $\sim$7~km~s$^{-1}$, which do not appear in the observed diagram. 

The SO line profile and PV diagram of Model 2 are presented in Fig. 1d and Fig. 3f,
respectively. Although the line profile made from Model 2 still does not have a single peak at the systemic velocity like the observations, its peaks appear close to the systemic velocity. 
The diagram made from Model 2 is quite similar to the observations, including the overall velocity gradient. On the other hand, the model diagram has two peaks at  $\sim$5.2~km~s$^{-1}$ and $\sim$6.7~km~s$^{-1}$ in LSR velocity, while the two peaks in the observed diagram appear at slightly different LSR velocities ( $\sim$5.9~km~s$^{-1}$ and $\sim$6.5~km~s$^{-1}$) 
These results suggest that Model 2 seems to reproduce the observations better than Model 1 overall.

The key feature to reproduce the observations is an infall velocity slower than
the free-fall velocity yielded by the central star. In our calculation described above, the free-fall velocity was estimated from the dynamical mass of $0.3~M_{\sun}$, whereas our conclusion is still valid even if we use the smaller mass of the central protostar, $0.2~M_{\sun}$, estimated by \citet{tobin12} \citep[see also][]{sakai14}. Although we also tried
different values for other parameters, such as 
the density profile (absolute value of the density and the power-law index) and the power-law index) of the model envelope, the power-law index of the temperature profile, and
the specific angular momentum of the infalling gas,
they were not essential to reproduce
the observed characteristics.
The deceleration of the infall velocity may be due to magnetic fields. 
For the region at radii $\geq$250~AU, magnetic braking may be a possible mechanism for
the  deceleration of the infall velocity.
In fact, in previous MHD simulations of star formation processes, it was demonstrated
that the accretion flow perpendicular to the magnetic field lines is
considerably slower than the free-fall velocity \citep{machida11,tomida13}.
In contrast to the region at radii $\geq$250~AU, it is not very clear how the infalling velocity
in the hot ring region can be decelerated. 
One possible mechanism to decelerate the infall velocity is the so-called magnetic wall \citep{li96,tassis05a,tassis05b}, where magnetic fields are transported from the inner region
by non-ideal MHD effects. In this region, the enhanced magnetic pressure can
decelerate the accretion flow and form a shock. 
Another possible mechanism to decelerate the infall velocity in the hot ring is 
spiral arms possibly formed around the Keplerian disk by transportation of angular momentum
from the disk.
Many hydrodynamic simulations show that spiral arms with large specific angular momentum spread into outer infall regions\citep[e.g.][]{saigo08,tsukamoto14}. Such spiral arms cause shocks by interactions with supersonic infalling gas.

\section{Conclusions}
L1527 IRS, a solar-type protostar surrounded by an infalling and rotating envelope, has been observed with ALMA in C$^{18}$O $2-1$, SO $6_5-5_4$ and 1.3 mm dust continuum. Our conclusions are summarized as follows;

\begin{enumerate}
\item
As previously shown, a flattened envelope was detected in C$^{18}$O around L1527 IRS. The emission shows strong redshifted absorption against the bright continuum emission associated with the protostar, strongly suggesting infall motions in the C$^{18}$O envelope. The C$^{18}$O envelope also shows differential rotation. The rotation velocity is mostly proportional to $1/r$, where $r$ is the radius, while the rotation profile at the innermost radius ($\sim$54~AU) may be shallower than $1/r$, suggestive of formation of a Keplerian disk around the central protostar of 
$\sim0.3~M_{\sun}$ in dynamical mass. 
\item
Compact SO emission arising from the inner part of the C$^{18}$O envelope was detected. The emission shows rotation in the same direction as the C$^{18}$O envelope. The rotation is, however, rigid-body like which is very different from the differential rotation shown by C$^{18}$O. The rigid-body like rotation can be naturally explained when the SO emission has a ring like structure, where the SO fractional abundance is locally enhanced due to higher dust temperature associated with a shock. The outer radius of the ring is estimated to be $\sim$120~AU, where SO molecules are locally enhanced due to a shock, while its inner radius is estimated to be $\sim$70~AU, where SO molecules would be depleted to dust grains again. 
\item
In order to explain the line profiles and the position-velocity (PV) diagrams of C$^{18}$O and SO observed, simple models composed of an infalling envelope surrounding a Keplerian disk of 54~AU in radius orbiting a star of 0.3~M$_{\sun}$ are examined. Models also include an SO ring region in the envelope, within which SO emission is detectable due to its fractional abundance enhancement, to reproduce the rigid-like velocity gradient of the SO emission. 
It is found that when the infall velocity in
the model is the same as the free-fall velocity yielded by a star of 0.3~M$_{\sun}$,
characteristic features of the observed line profiles and PV diagrams are not reproduced well.
In order to reproduce them, infall velocities have to be reduced by 1/2 at radii more than 250~AU
and by 1/5 within 120~AU except for the Keplerian disk where there is no infall.
\end{enumerate}

Although the mass of the protostar is one of the most crucial physical parameters to understand the formation and evolution of the protostar, it has been previously estimated from infall velocities or accretion luminosities with various assumptions. 
Detection of Kepler motions around protostars allows us to derive the dynamical masses of the protostars, which would be the only mean to directly estimate masses of protostars without any assumptions. The dynamical mass can be directly compared with various physical parameters, including the infall velocity or the accretion luminosity, and such comparison may shed a new light on the study of star formation.

\acknowledgments
This paper makes use of the following ALMA data:
ADS/JAO.ALMA\#2011.0.00210.S. . ALMA is a partnership of ESO (representing
its member states), NSF (USA) and NINS (Japan), together with NRC
(Canada) and NSC and ASIAA (Taiwan), in cooperation with the Republic of
Chile. The Joint ALMA Observatory is operated by ESO, AUI/NRAO and NAOJ.
We thanks all the ALMA staff making our observations successful. We also thank Takayuki Muto 
and Takaaki Matsumoto who provides us with fruitful comments on the manuscript.
S. K. is supported by the Subaru Telescope Internship Program.
K. T. is supported by JSPS Research Fellowship for Young Scientists.
S. T. acknowledges a grant from the National Science Council of Taiwan 
(NSC 102-2119-M-001-012-MY3) in support of this work.
This study is supported by Grants-in-Aid from the Ministry of Education, 
Culture, Sports, Science, and Technologies (23540266, 23103004 (Y. A)).




{\it Facilities:} \facility{ALMA}.



\appendix

\section{Statistical Equilibrium Calculations for SO emission}
In order to investigate the physical condition in the SO emission region, we performed
statistical equilibrium calculations of the three SO lines, which were observed in L1527 with ALMA (see Table~2), based on the large velocity gradient (LVG) model\citep{goldreich74}. For the calculations, all the three SO lines were resampled with the same velocity resolution (0.28~km~s$^{-1}$) and also convolved with the same beam size ($0\farcs96 \times 0\farcs73$). The brightness temperature of each SO line was measured at the systemic velocity within the measurement area of 0$\farcs75 \times 0\farcs75$ centered at the protostellar position. 
Rotational energy levels, statistical weights, Einstein A-Coefficients, and line frequencies of SO are
taken from the LAMDA database\citep{schoier05}. Collisional transition rates of SO below 50~K are
taken from \citet{lique07}, and those above 60~K are taken from \citet{lique06}.
The escape probability $\beta$ in the case of a static, spherically symmetric and homogeneous medium\citep{osterbrock06} is adopted. 
The rotational energy levels included in the calculations are up to the 91st rotational energy level ($E_u=$ 972~K) in the case of above 60~K, and the 31st energy level ($E_u=$ 127~K) in the case of below 50~K.
These energy levels are high enough to discuss the physical conditions of molecular gas in low-mass
protostellar envelopes. We confirmed that our calculations provide the same results as those from the RADEX code\citep{tak07}.

There are three free parameters for the LVG calculations, i.e., the kinetic temperature, the number density of H$_2$, and $\frac{X}{dV/dr}$, where $X$ is the fractional abundance of the SO molecule and $dV/dr$ is the velocity gradient. Brightness temperatures of each SO line are calculated with various sets of the three parameters, and are compared with the observed brightness temperatures with a $\chi^2$ test to find the best solution explaining the observations. When the calculated brightness temperatures are compared with the observed values, it is important to consider the filling factor that is the fraction of the emission size to the measurement area. 
In oder to estimate the filling factor, we consider which part of the edge-on ring can be observed at the systemic velocity within the measurement area of 0$\farcs75 \times 0\farcs75$ (105~AU $\times$ 105 AU) centered at the protostellar position. The edge-on ring is observed within the measurement area as a rectangular, and the filling factor is given as $S_{\rm V}\times S_{\rm H}/(105~{\rm AU})^2$, where $S_{\rm V}$ and $S_{\rm H}$ are the size of the rectangular along the vertical and horizontal directions, respectively. $S_{\rm V}$ and $S_{\rm H}$ are estimated as follows;
\begin{enumerate}
\item
It is considered that $S_{\rm V}$ is determined by the scale hight of the ring $h(T)$, which is expressed as  
\begin{equation}
h(T)=0.05 \times \left(\frac{T}{300~{\rm K}}\right)^{0.5 }\left(\frac{R}{1~{\rm AU}}\right)^{1.5}\left(\frac{M}{M_\sun}\right)^{-0.5}~{\rm AU}
\end{equation}
where $T$ is the temperature, $R$ is the distance from the central star, and $M$ is the mass of the central star. When $R=$120~AU and $M=0.3~M_\sun$, 
\begin{equation}
S_{\rm V}=2h(T)=240\left(\frac{T}{300~{\rm K}}\right)^{0.5 }~{\rm AU}
\end{equation}
\item
$S_{\rm H}$ is estimated by considering the part of the edge-on ring with its line-of-sight velocity less than 0.14~km~s$^{-1}$ with respect to the systemic velocity because emission within $\pm$0.14~km~s$^{-1})$ with respect to the systemic velocity is detected at the central channel with a velocity resolution of 0.28~km~s$^{-1}$. Since the ring is considered to rotates at 1~km~s$^{-1}$ speed at the radius of 120~AU, $S_{\rm H}$ is estimated to be $\sim$34~AU.
\end{enumerate}

Fig.~7 summarizes the results of the  $\chi^2$ test. Each panel shows a result with a different value of $\frac{X}{dV/dr}$.
We consider that parameter sets with $\chi^2<10$ can reproduce the observations very well.
The area with $\chi^2<10$ appears at a temperature of $\sim$32~K at each panel, while it spreads over a wide range of the H$_2$ density. This indicates that the temperature can be determined well with the current LVG analysis, while it is difficult to constrain the H$_2$ density and $\frac{X}{dV/dr}$.
This is quite reasonable when the optical depth of the SO line is examined; the part where $\tau=1$ shown in each panel indicates that the best solutions with $\chi^2<10$ appear when SO is optically thick. Since SO is optically thick, it is impossible for us to estimate the H$_2$ density and the abundance of SO. With this LVG analysis, we consider that the kinetic temperature of the SO gas within the ring is $\sim$32~K.

\clearpage



\begin{figure}
\epsscale{1.0}
\plotone{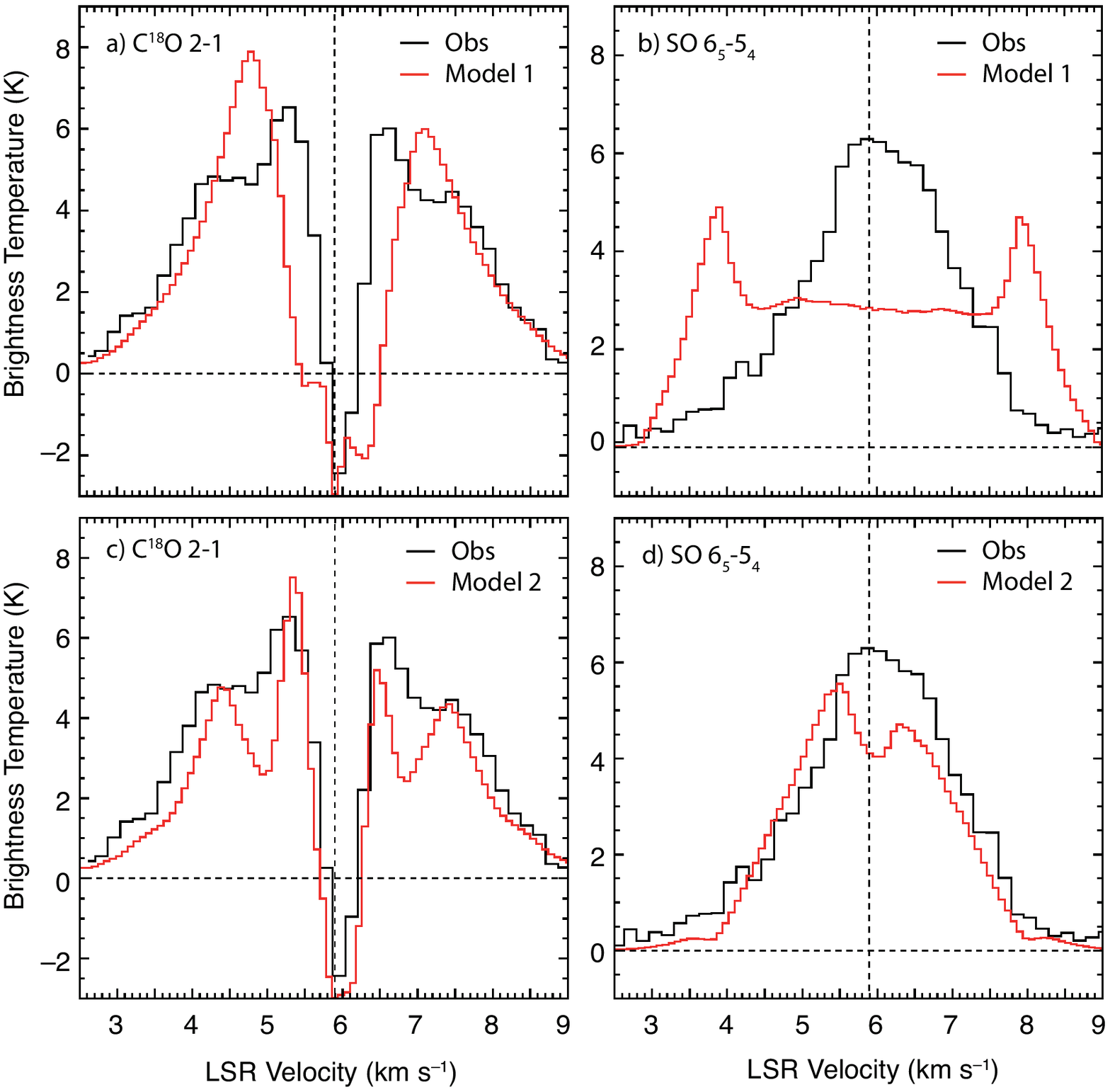}
\caption{Line profiles of C$^{18}$O $2-1$ and SO $6_5-5_4$ obtained  
at the central star position within an area of $0\farcs75 \times 0\farcs75$. The vertical dashed line in each 
panel shows the systemic velocity of 5.9~km~s$^{-1}$. Observed line profiles are shown in black lines, while those calculated by models are shown in red lines. C$^{18}$O line profiles are presented in Panel (a) and (c), while SO line profiles are presented in Panel (b) and (d).\label{fig1}}
\end{figure}

\clearpage


\begin{figure}
\epsscale{1.0}
\plotone{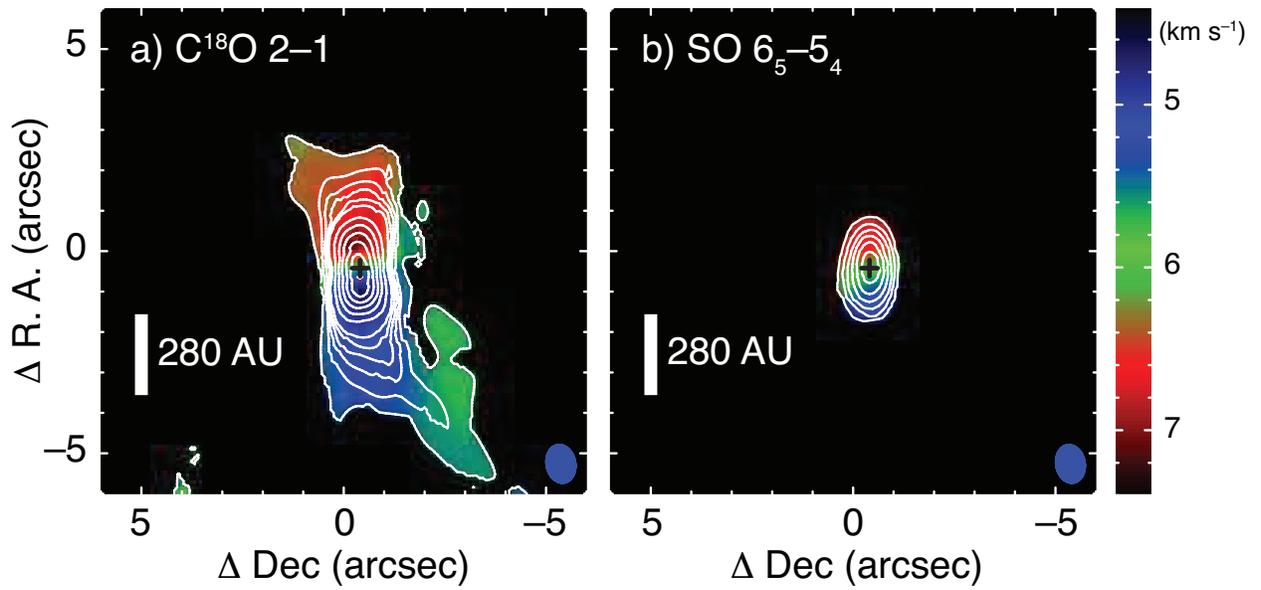}
\caption{Integrated intensity (moment~0) shown in contours and intensity-weighted mean velocity (moment~1) maps shown in color of C$^{18}$O $2-1$ (left-hand side) and SO $6_5-5_4$ (right-hand-side). Contours are drawn from 3~$\sigma$ to 15~$\sigma$ in steps of 3~$\sigma$ and in steps of 5~$\sigma$ for more than 15~$\sigma$.The cross and the blue ellipse at the bottom-right corner in each panel show the position of the central protostar L1527 IRS and the synthesized beam.\label{fig2}}
\end{figure}


\begin{figure}
\epsscale{0.7}
\plotone{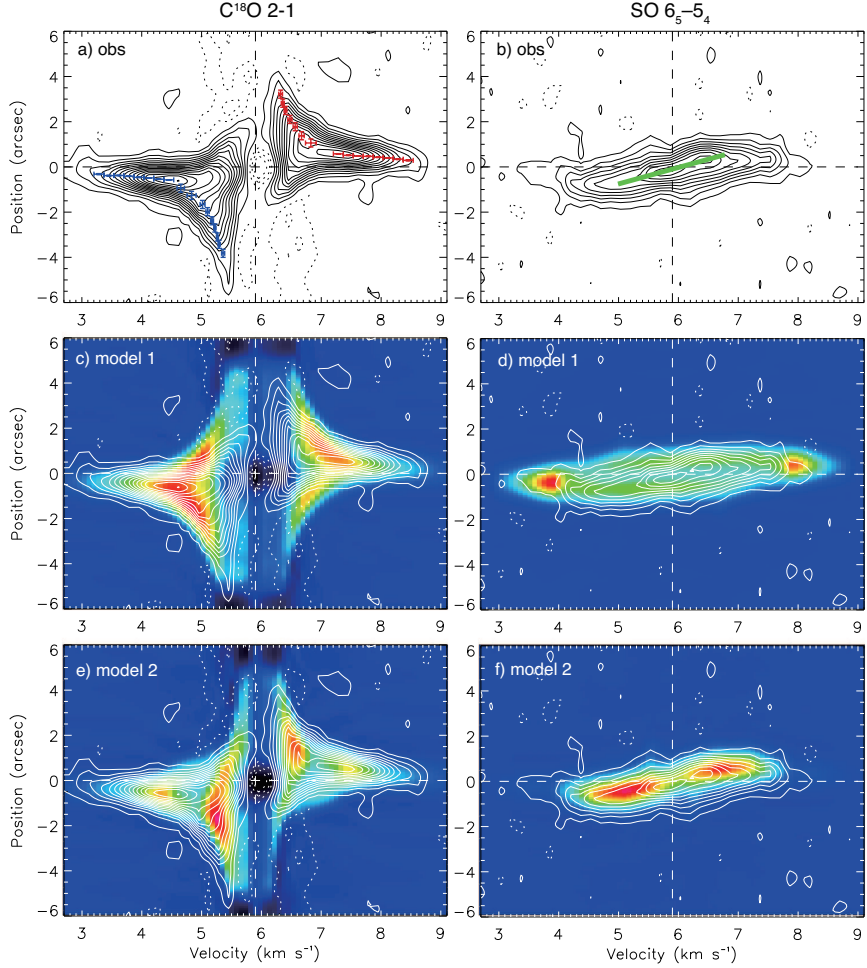}
\caption{Position-Velocity (PV) diagrams along the line from the north to south passing through the position of the central star. Left-hand panels show diagrams of C$^{18}$O while right-hand panels
show those of SO. Diagrams obtained from the observations are shown in black or white contours, while diagrams calculated from models are shown in color. 
The solid contours are drawn from $+3~\sigma$ with a 3~$\sigma$ step, while the dotted contours are drawn in the same way but from $-3~\sigma$. The vertical and horizontal dashed lines in each panel show the systemic velocity and the position of the central star, respectively.  (a) Diagram obtained from the observations. Blue and red marks show representative data points in the diagram. (b) Diagram obtained from the observations. (c) Model diagram obtained from Model 1 is compared with
the observed diagram shown in (a).
(d) Model diagram obtained from Model 1 is compared with the observed diagram shown in (b).
(e) Same as (c) but model diagram is obtained from Model 2. (f) Same as  (d) but the model diagram is obtained from Model 2.\label{fig3}}
\end{figure}

\begin{figure}
\epsscale{1.0}
\plotone{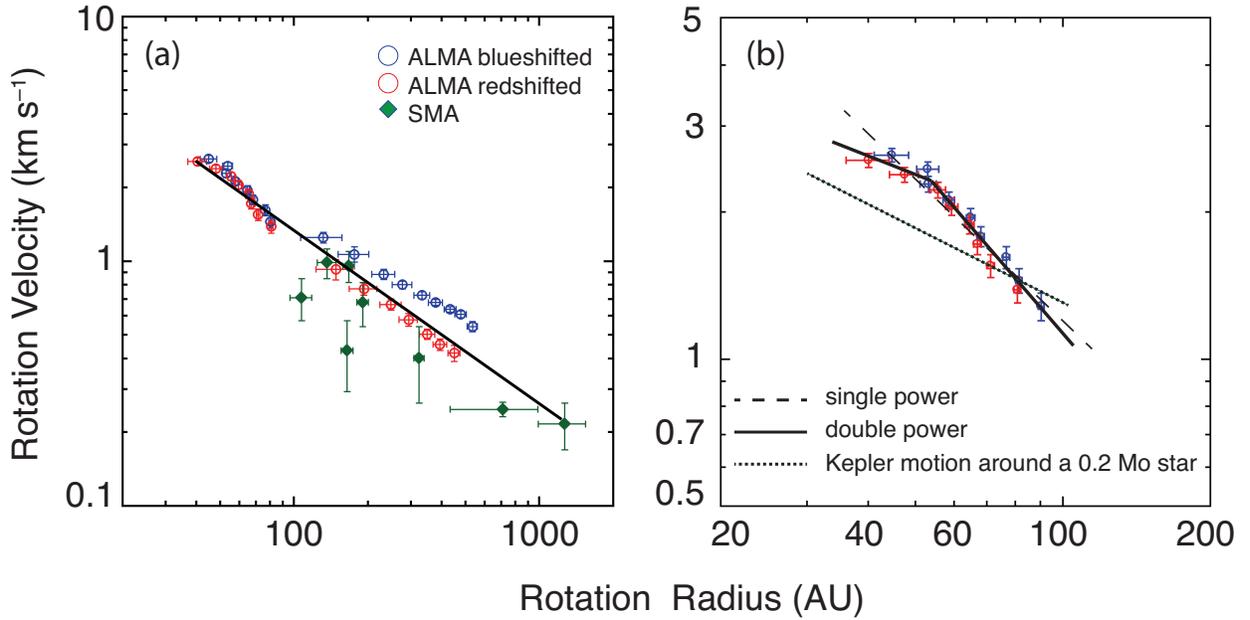}
\caption{Rotation profiles of the C$^{18}$O flattened envelope. (a) Blue and red marks show the ALMA data points measured respectively at blueshifted and redshifted velocities in the PV diagram shown in Fig.~3a. For comparison, the SMA measurements\citep{yen13} are also plotted in green marks. Note that the systemic velocity adopted in \citet{yen13} was 5.7~km~s$^{-1}$, while we adopted 5.9~km~s$^{-1}$ for both the ALMA and the SMA measurements. The solid line shows the result of the least-square fitting to all the data points including both the SMA and the ALMA measurements. (b) Only the ALMA data points within 100~AU in rotation radius are shown. The dashed line shows the least-square fitting with a single power-law, while the solid line shows the fitting with two power-laws. The dotted line shows the rotation profile of the Kepler rotation around a 0.2$M_{\sun}$ star.\label{fig4}}
\end{figure}

\begin{figure}
\epsscale{1.0}
\plotone{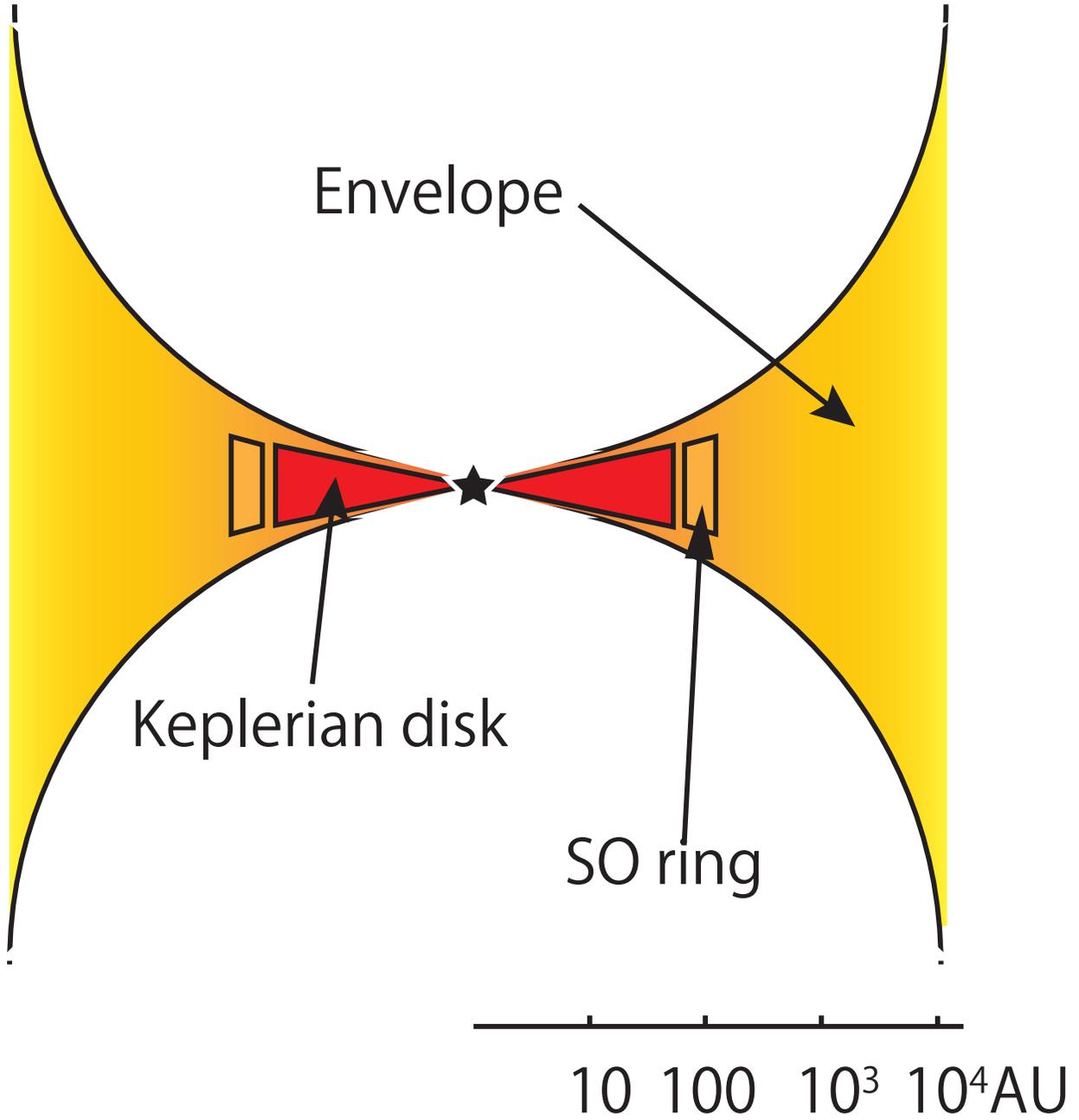}
\caption{Schematic view of the cross section of the model. The distance from the central star
in AU is also shown at the bottom.}
\end{figure}

\begin{figure}
\epsscale{1.0}
\plotone{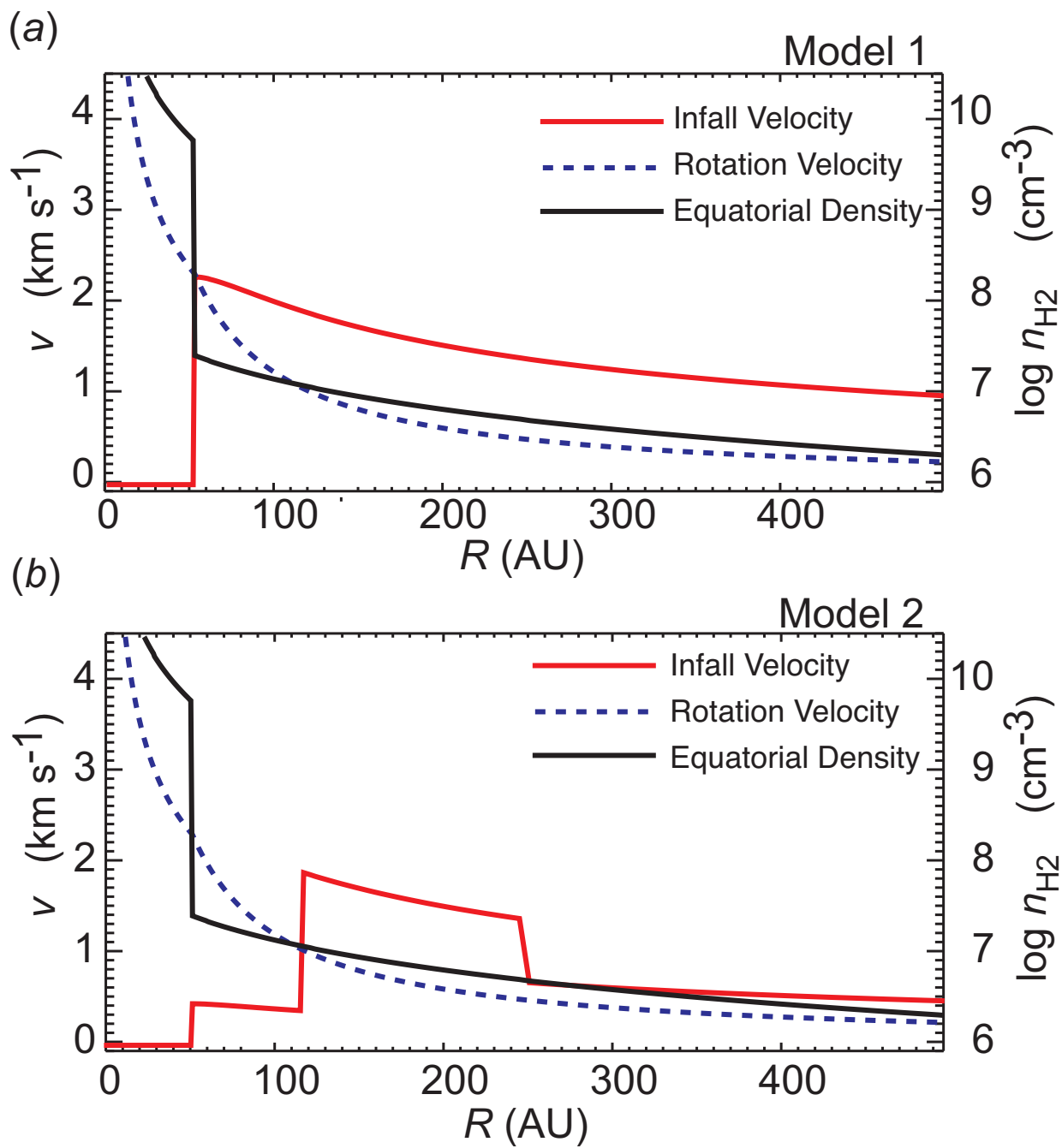}
\caption{Velocity and density radial profiles for (a) Model~1 and (b) Model~2. 
Red solid and dashed curves show the infall velocity and the rotation
velocity distributions on the equatorial plane, respectively, while dashed curves show the
density distribution.}
\end{figure}

\begin{figure}
\epsscale{1.0}
\plotone{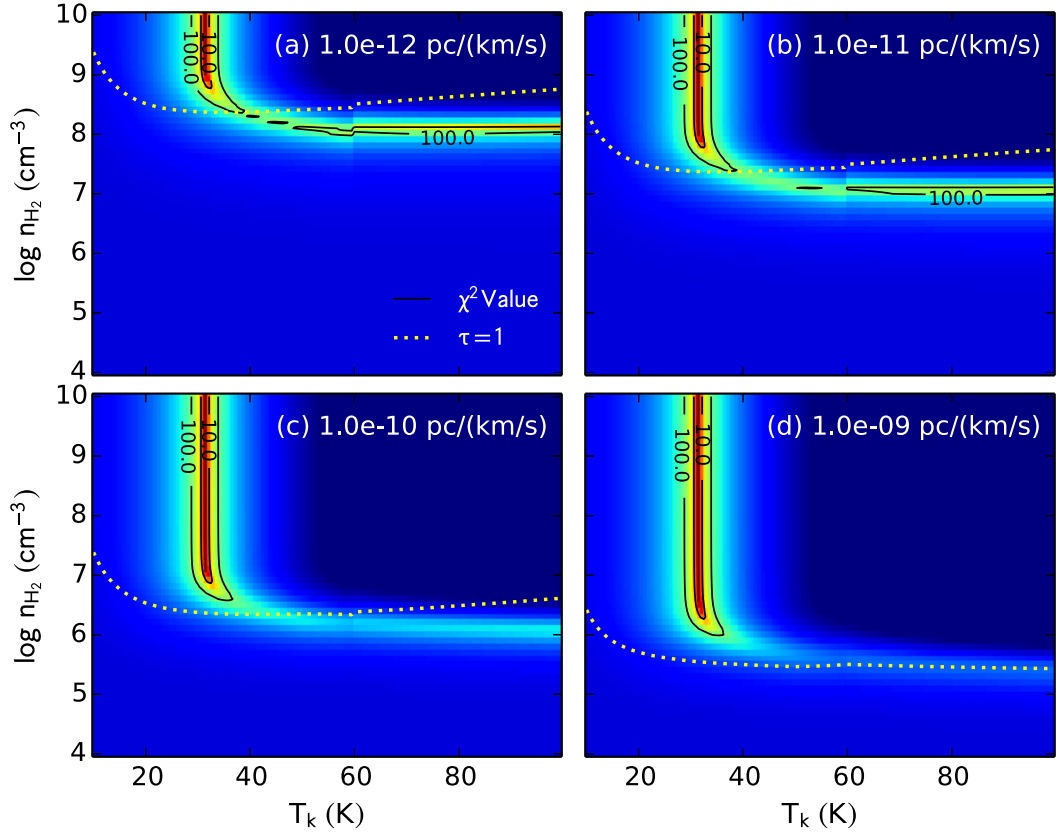}
\caption{$\chi^2$ distribution in $T_k$ - $n_{{\rm H}_2}$ space with different $\frac{X}{dV/dr}$ values, which are shown at the top-right conner in each panel. Distribution of $\chi^2$ is shown in color and black contours, while the part where the optical depth of SO is 1 is shown in dotted yellow contours.} 
\end{figure}

\clearpage
\begin{table}[htbp]
\begin{center}
\caption{Summary of the Observational Parameters}
\scalebox{0.75}{
\begin{tabular}{lccc}
\hline \hline
 & SO $6_5$--$5_4$ & C$^{18}$O 2--1 & continuum\\ 
Interferometer and date & \multicolumn{3}{c}{ALMA, 2012.Aug.26}\\
\hline
Target & \multicolumn{3}{c}{L1527}\\
Coordinate center & \multicolumn{3}{c}{R.A. (J2000)=4$^{\rm h}$39$^{\rm m}$53$^{\rm s}\!\!$.9000}\\
 & \multicolumn{3}{c}{Dec. (J2000)=26$^{\circ }03\arcmin 10\farcs 000$}\\
Frequency &219.9494 GHz&219.5603 GHz&225.4336 GHz \\
Primary beam &$28\farcs 6$&$28\farcs 6$&$27\farcs 9$ \\
Projected baseline length & \multicolumn{3}{c}{18.0 -- 372.5 m}\\
Synthesized beam (P.A.) &$0\farcs 96\times 0\farcs 73\ (+10^{\circ })$&$0\farcs 96\times 0\farcs 73\ (+11^{\circ })$&$0\farcs 93\times 0\farcs 70\ (+12^{\circ })$\\
Velocity resolution & 0.17 km$\,$s$^{-1}$ & 0.17 km$\,$s$^{-1}$ & 94 MHz\\
Noise level (no emission)& 8.0 mJy$\,$beam$^{-1}$ & 6.6 mJy$\,$beam$^{-1}$& 0.53 mJy$\,$beam$^{-1}$\\
Noise level (detected channel)& 9.5 mJy$\,$beam$^{-1}$ & 8.0 mJy$\,$beam$^{-1}$& ---\\
Passband calibrator & \multicolumn{3}{c}{J0522-364}\\
Flux calibrator & \multicolumn{3}{c}{Callisto}\\
Gain calibrator & \multicolumn{3}{c}{J0510+180}\\
\hline
\end{tabular}
}
\end{center}
\end{table}

\clearpage
\begin{table}
\begin{center}
\caption{Summary of SO lines for LVG calculations.\label{tbl-2}}
\begin{tabular}{lccc}
\hline \hline
& SO $6_5$--$5_4$ & SO $7_6$--$6_5$ & SO $7_8$--$6_7$\\
\hline
Frequency (GHz) & 219.94944 & 261.843721 & 340.714155\\
E$_{\rm u}$ (K) & 34.98 & 47.55 & 81.24\\
Observed Brightness Temperature (K) & 6.16$\pm$0.30 & 6.00$\pm$0.15 & 5.64$\pm$0.11\\
Reference & 1 & 2\tablenotemark{a} & 2\tablenotemark{a}\\
\hline
\end{tabular}
\end{center}
\tablenotetext{a}{For the lines observed by \citet{sakai14}, the calibrated data was downloaded from the ALMA archive, and clean maps were made by ourselves.}
\tablerefs{
(1) This work; (2) Sakai et al. (2014)}
\end{table}




\clearpage




\end{document}